\documentclass[aps,prl,amsmath,amssymb,groupedaddress,superscriptaddress,floatfix,twocolumn,showkeys,amsart]{revtex4-1}
\usepackage{graphicx} 
\usepackage[caption=false]{subfig}
\usepackage{float}
\usepackage{xcolor}

\usepackage{comment}
 
\begin{document}
\title{Understanding the Quality Factor of Mass-Loaded Tensioned Resonators}
\begin{abstract}
   Mechanical resonators featuring large tensile stress have enabled a range of experiments in quantum optomechanics and novel sensing. Many sensing applications require functionalizing tensioned resonators by appending additional mass to them. However, this may dramatically change the resonator mode quality factor, and hence its sensitivity. In this work, we investigate the effect of the crossover from no mass load to a large mass load on the mode shape and quality factor of a tensioned resonator. We show through an analytical model and finite element analysis that as the load mass increases, surprisingly, the resonator mode shape becomes independent of the exact load mass, and therefore, its quality factor saturates. We validate this saturation effect experimentally by measuring the quality factor of a silicon nitride trampoline resonator while varying the load mass in a controlled manner.
\end{abstract}
\author{R. Shaniv, S. Kumar Keshava, C. Reetz, and C. A. Regal}
\date{\today}
\maketitle

\par Micromechanical and nanomechanical resonators have been a fruitful topic of research for several decades. They have shown promise for exploring quantum physics~\cite{aspelmeyer2014cavity}, in studies ranging from quantum transduction protocols~\cite{andrews2014bidirectional, forsch2020microwave, mirhosseini2020superconducting}, creation of quantum memories~\cite{wallucks2020quantum, kharel2018ultra,maccabe2020nano}, to the generation of squeezed light~\cite{purdy2013strong, safavi2013squeezed,brooks2012non}, and for precision sensing applications, such as magnetic force detection~\cite{rugar2004single, scozzaro2016magnetic, fischer2019spin}, nanoscale imaging~\cite{degen2009nanoscale, halg2021membrane}, accelerometer design~\cite{krause2012high, zhou2021broadband}, and mass sensing~\cite{yang2006zeptogram, ekinci2004ultrasensitive, lavrik2003femtogram}.

\par An adequate sensor requires both appreciable coupling to the physical parameter it is designed to sense and as low as possible environmental noise. An important figure of merit that quantifies the sensitivity of a resonator normal mode is the quality factor, denoted hereafter as $Q$. The $Q$ depends on environmental isolation and is a measure of the number of coherent oscillations a resonator mode can provide before losing a significant portion of its energy to the environment~\cite{taylorclassical, aspelmeyer2014cavity}. Therefore, resonator modes with utmost $Q$ values are ideal for sensing.

\par To couple a sensor to a quantity of interest, it is often necessary to functionalize it with a coupling agent, resulting in an additional resonator mass, e.g.~magnetic particles for magnetic force sensing~\cite{rugar2004single, fischer2019spin} or test masses for acceleration detection~\cite{krause2012high,zhou2021broadband}. Certain applications, such as those that aim to measure gravitational forces, require appending a mass that is much larger than the resonator mass~\cite{liu2021gravitational, schmole2016micromechanical, pratt2021nanoscale}. In general, specific interest is devoted to modes in which the mass moves appreciably. For a large mass, typically, a single mode has a significant motion in the load, and it is usually the lowest frequency or fundamental mode. Although essential for sensing, the addition of a load mass might lower the resonator's $Q$, and degrade its performance as a sensor. It is therefore of interest to understand how the $Q$ of a resonator mode is affected by mass loading.

\par The $Q$ of a resonator mode is defined as $Q = 2\pi\frac{W}{\Delta W}$, where $W$ is the energy stored in the mode, and $\Delta W$ is the energy loss per oscillatory cycle~\cite{taylorclassical}. In general, $\Delta W$ is a sum of the contributions from multiple individual loss mechanisms~\cite{schmid2016fundamentals}. One way to improve the $Q$ of a resonator is by increasing the stored energy without a similar increase in the dissipation. This can be achieved by using a tensioned or high-stress film as the resonator and it is the result of a phenomenon called dissipation dilution~\cite{fedorov2019generalized,tsaturyan2017ultracoherent, ghadimi2018elastic}. In addition to increased $Q$, high-stress also extends the mode spectrum of a resonator to higher frequencies for a given resonator length scale. High frequency is of interest for certain sensing protocols, while being out of reach for similar-dimension non-tensioned sensor geometries such as cantilevers.

\par  Low optical absorption combined with high-stress obtained in fabrication, make silicon nitride (SiN) a natural material for the design of high $Q$ resonators~\cite{zwickl2008high, yuan2015silicon,serra2016microfabrication}. In tensioned SiN resonators, typically, the $Q$ of modes are limited by two loss mechanisms --- bending loss, which is the energy lost due to the bending of the mode, and radiation loss, which is the energy lost from the mode via acoustic radiation into the surroundings. This allows us to define individual quality factors, $Q_{\mathrm{bend}}$ and $Q_{\mathrm{rad}}$, which are associated with bending loss and radiation loss respectively. Therefore, we can write that $\frac{1}{Q}=\frac{1}{Q_{\mathrm{bend}}} + \frac{1}{Q_{\mathrm{rad}}}$.  $Q_{\mathrm{bend}}$ can be significantly improved by tailoring the mode-shape to reduce bending loss at the mode edge, commonly referred to as clamping loss~\cite{reetz2019analysis,tsaturyan2017ultracoherent, bereyhi2019clamp, reinhardt2016ultralow, fedorov2020fractal, norte2016mechanical, ghadimi2018elastic, ghadimi2017radiation}. $Q_{\mathrm{rad}}$ can be improved by resonator patterning, substrate engineering, and strategic device mounting~\cite{reetz2019analysis,tsaturyan2017ultracoherent, yu2014phononic, weaver2016nested, nortethesis, dalthesis, reinhardtthesis, borrielli2016control,li2016suppression, ghadimi2017radiation}. These techniques have led to resonators with modes limited by $Q_{\mathrm{bend}}$.

\par In this work, we study the effect of a localized mass load on the $Q_{\mathrm{bend}}$ of highly tensioned resonators. We show through analytical calculations and finite element analysis (FEA) simulations that as the load mass increases, the modes of the resonator change in frequency and displacement shape. Further, we show that for a large enough mass, each mode displacement function becomes independent of the mass, which we claim leads to mass-independent $Q_{\mathrm{bend}}$. We refer to this phenomenon as $Q$ \textit{mass saturation}. We validate this saturation experimentally by measuring the $Q$ of a tensioned SiN trampoline resonator~\cite{norte2016mechanical,reinhardt2016ultralow} as a function of the load mass. We use magnetic grains for the load mass, and we vary the mass by sequentially stacking the grains using their mutual magnetic attraction to avoid varying the amount of lossy adhesive as mass was added. This allows us to compare $Q$ measurements for different load masses on a single device.  
\par In order to obtain an expression for the $Q_{\mathrm{bend}}$ of a mode, we assume pure out-of-plane mode displacement $u\left(x,y\right)$, where $x$ and $y$ are in-plane resonator coordinates, as well as small displacement, $\left|u\left(x,y\right)\right|\ll h$, where $h$ is the resonator thickness. In addition, we assume large in-plane tensile stress $\sigma_{0}$, such that the speed of sound along the resonator is approximately proportional to $\sqrt{\sigma_{0}}$. Then we can write $W\approx\int_{V}\frac{\sigma_{0}} {2}\left[\left(\frac{\partial u}{\partial x}\right)^{2}+\left(\frac{\partial u}{\partial y}\right)^{2}\right]dV$ and $\Delta W_{\mathrm{bend}} \approx \int_{V}\frac{\pi E_{2} }{(1-\nu^{2})}z^2\left[\frac{\partial^{2} u}{\partial x^{2}}+\frac{\partial^{2} u}{\partial y^{2}}\right]^{2} dV$, where $E_2$ is the imaginary part of the Young modulus, $\nu$ is the Poisson ratio, $z$ is the resonator coordinate along its thickness and $V$ is the volume of the resonator~\cite{tsaturyan2017ultracoherent, yu2012control}. This leads to:
\begin{equation} \label{Geometric_Q}
        Q_{\mathrm{bend}}\left(u\right) =  \frac{\int_{V}\sigma_0\left[ \left(\frac{\partial u}{\partial x}\right)^{2}+\left(\frac{\partial u}{\partial y}\right)^{2}\right] dV}{\int_{V}\frac{E_{2} }{(1-\nu^{2})}z^2\left[\frac{\partial^{2} u}{\partial x^{2}}+\frac{\partial^{2} u}{\partial y^{2}}\right]^{2} dV}
\end{equation}
Eq.~\ref{Geometric_Q} affirms that if the mode displacement function $u(x,y)$ stops changing, then $Q_{\mathrm{bend}}$ saturates. 


\begin{figure}[t]
\centering
\includegraphics[width=0.50\textwidth]{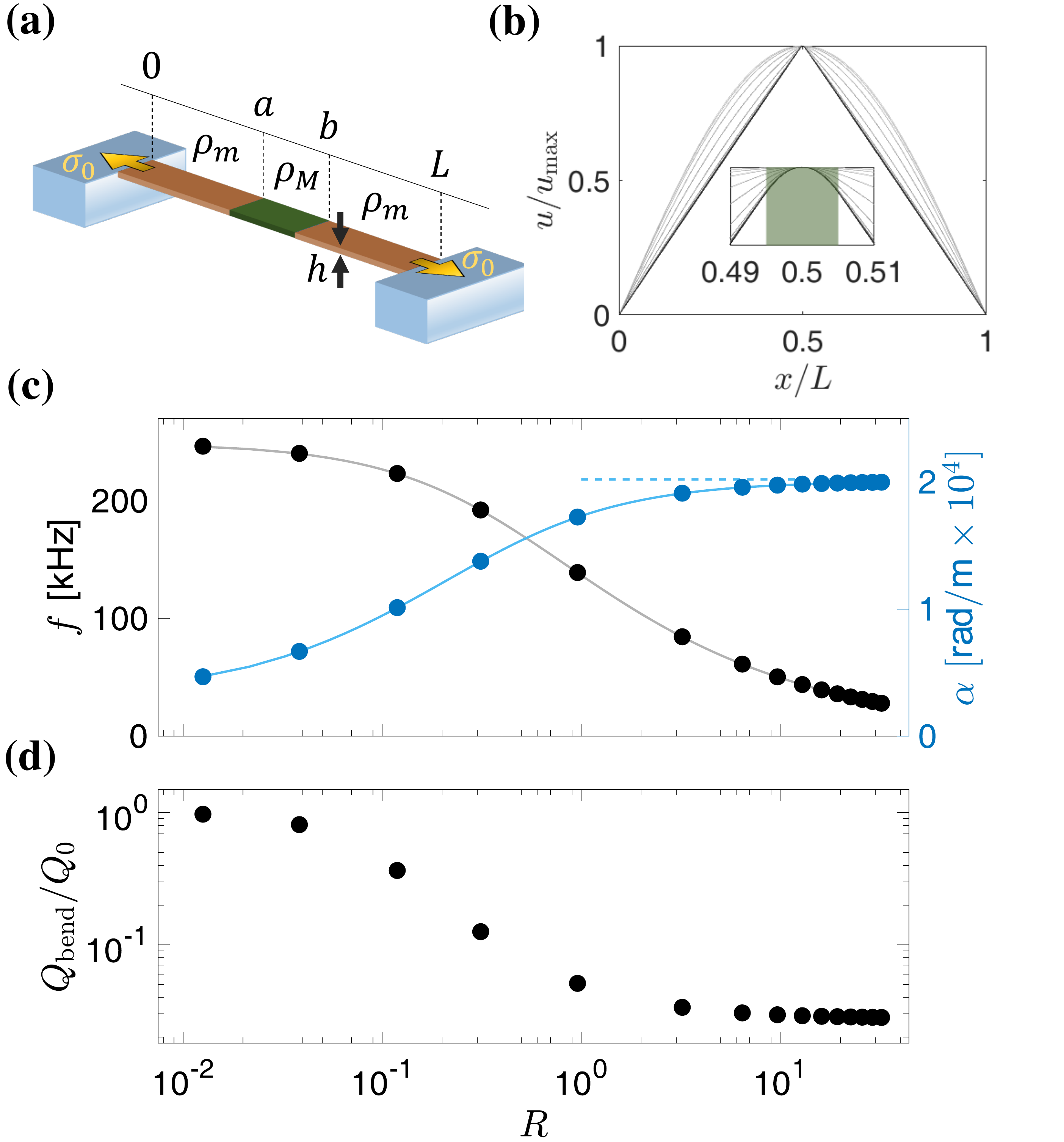}
\caption{\textbf{Mass loaded string theoretical model.}\newline
\textbf{(a)} String parameters: A string of length $L$, thickness $h$, with tensile stress $\sigma_{0}$ is fixed at points $x=0$ and $x=L$, where $x$ is the coordinate along the length of the string, and has linear density $\rho_{m}$ everywhere except for the region $a\le x\le b$, where the density is $\rho_{M}+\rho_{m}$. \textbf{(b)} Mode shape visualization: FEA simulations show the mass loaded string fundamental mode displacement normalized to its maximum displacement (lighter color corresponds to lighter mass load) plotted against the spatial coordinate normalized to the string length. The inset shows a zoomed image of the mode shape around the mass loaded region (shaded green). \textbf{(c)} String frequency and inner region wavenumber: FEA simulated (solid circles) and analytically calculated (solid lines) fundamental mode frequency (black) and inner region wavenumber (blue) of a string are shown as a function of $R$, the ratio between the mass of the load and the total mass of the unloaded resonator. FEA simulation points correspond to the different mode shapes in subfigure  \textbf{(b)}. The analytically calculated large mass limit inner region wavenumber $\alpha_{\mathrm{lim}}$ is indicated (dashed blue line). \textbf{(d)} String quality factor: The plot shows the fundamental mode $Q_{\mathrm{bend}}$ normalized to $Q_{0}$, the $Q_{\mathrm{bend}}$ of an unloaded resonator, as a function of the mass ratio $R$. For this plot, to focus on the effect of the mass load, $Q_{\mathrm{bend}}$ was calculated without the mechanical clamping loss contribution from the edges of the mode. Subfigures \textbf{(c)} and \textbf{(d)} share the same horizontal coordinate, and results shown correspond to specific choice of parameters (supplementary material).} 
\label{Theoretical_model}
\end{figure}
\par We can elucidate this $Q$ saturation and its origin through an example model. We study the fundamental mode of a highly pre-tensioned suspended beam, which is referred to as a string. (A second example for a tensioned, mass loaded circular membrane can be found in the supplementary material.) The string resonator is fixed at both ends, with displacement $u\left(x\right)$ in the vertical axis, where $x$ is the coordinate along the resonator (Fig.~\ref{Theoretical_model}a). 
Here $u\left(x\right)$ satisfies the string equation~\cite{schmid2016fundamentals}:
\begin{equation}  \label{Full_string_equation}
    \frac{E h^{2}}{12\sigma_{0}}\frac{d^{4}u}{dx^{4}}-\frac{d^{2}u}{dx^{2}}-\frac{\rho(x) (2\pi f)^{2}}{\sigma_{0}}u = 0,
\end{equation}
for $0\leq x\leq L$, where $L$ is the string length. Here, $E$, $h$, $\sigma_{0}$, are the string's Young modulus, thickness and the tensile stress respectively, and $f$ is the mode frequency. The string width $w$ is absent from the equation because the string has a uniform width. The position-dependent density $\rho\left(x\right)$ accounts for possible increased density at some region $a \leq x \leq b$, for $0 < a < b < L$. We denote the density at $0\leq x < a$ and $b < x \leq L$, subsequently referred to as the outer regions, by $\rho_{m}$, and the density at $a \leq x \leq b$, the inner region, by $\rho_{m} + \rho_{M}$, such that $\rho_{M}$ is the added load density. The pre-stress $\sigma_{0}$ is assumed to be large, satisfying the length scale inequality $l = 2\pi\sqrt{\frac{E h^{2}}{12\sigma_{0}}}\ll L$.

\par Given this high pre-stress condition the frequency of the string's fundamental mode can be well approximated by neglecting the fourth derivative term in Eq.~\ref{Full_string_equation} ~\cite{schmid2016fundamentals}. The solution then reads:
\begin{flalign} \label{high_mass_load_solution}
   &u\left(x\right) = \nonumber\\ 
   &\begin{cases}
    A_{1}\sin\left(\frac{\alpha}{\sqrt{1+\chi}} x\right) & 0\leq x \leq a\\
    A_{2}\sin\left(\alpha x\right) + B_{2}\cos\left(\alpha x\right) & a\leq x \leq b \\
    A_{3}\sin\left(\frac{\alpha}{\sqrt{1+\chi}} \left(L-x\right)\right) & b\leq x \leq L
    \end{cases},
\end{flalign}
where we introduced the density ratio $\chi=\frac{\rho_{M}}{\rho_{m}}$ and the inner region wavenumber $\alpha=2\pi f\sqrt{\frac{\rho_{m}+\rho_{M}}{\sigma_{0}}}$, and imposed the fixed boundary conditions $u\left(0\right)=u\left(L\right)=0$. An equation for $\alpha$ can be found by using the continuity of $u\left(x\right)$ and its first derivative at $x=a$ and $x=b$ (supplementary material). We see excellent agreement between FEA simulations of $\alpha$ and $f$ for the fundamental mode of a string, and the corresponding numerically-evaluated analytical calculation, for different values of $R$ the resonator mass ratio (Fig.~\ref{Theoretical_model}c). We define the resonator mass ratio as $R=\frac{M_{\mathrm{load}}}{M_{\mathrm{unloaded}}}$ using the quantities $M_{\mathrm{load}}=hw\left(b-a\right)\left(\rho_{M}\right)$ and $M_{\mathrm{unloaded}}=hwL\rho_{m}$ corresponding to the load mass and the total unloaded resonator mass, respectively. This parameter was chosen to emphasize that we are investigating the crossover between no load ($R=0$) and a load that is significantly more massive than the entire resonator mass ($R\gg 1$). 

\par In addition to $\alpha$ and $f$, we used the results from the FEA simulations for the fundamental mode along with Eq.~\ref{Geometric_Q} to calculate $Q_{\mathrm{bend}}$.  Fig.~\ref{Theoretical_model}d shows the dimensionless geometric parameter $Q_{\mathrm{bend}}/Q_{\mathrm{0}}$, where $Q_{\mathrm{0}}$ is the $Q_{\mathrm{bend}}$ without a load mass, as a function of $R$. While calculating $Q_{\mathrm{bend}}$, in order to focus specifically on the effect of the load mass, we chose to exclude the mechanical clamping loss contribution from the edges of the mode. From the plot, we see that $Q_{\mathrm{bend}}/Q_{\mathrm{0}}$ becomes independent of the exact load mass for $R\gg 1$ and it saturates to a value that is lower than when $R$ is close to zero.  (Note the plot points corresponding to $R=0$ have been omitted from Fig.~\ref{Theoretical_model}c and Fig.~\ref{Theoretical_model}d because the horizontal axis is on a logarithmic scale.)   

\par Fig.~\ref{Theoretical_model}b evinces that the fundamental mode inner region wavenumber $\alpha$ saturates to a finite value, $\alpha_{\mathrm{lim}}$, as the mass grows larger (supplementary material). Simultaneously, the outer region wavenumber diminishes, and in the large mass limit the mass loaded string equation can be approximated as
\begin{equation} \label{high_mass_string_equation}
    \frac{l^{2}}{(2\pi)^{2}}\frac{d^{4}u}{dx^{4}}-\frac{d^{2}u}{dx^{2}}-K^{2}\left(x\right)u = 0,
\end{equation}
where
\begin{equation} \label{high_mass_string_equation_wave_number}
    K^{2}(x) \approx \begin{cases}
   0 & 0\le x < a ,\quad b < x \le L\\
\alpha_{\mathrm{lim}}^{2} & a\leq x \leq b 
\end{cases}  
\end{equation}
Because this equation is independent of the load mass, so is its solution (supplementary material). 

\par In the high mass load limit, the mode shape of the unloaded regions can be understood in two equivalent ways. The first is temporal: for an unloaded resonator, the fundamental mode period is given by $\sqrt{\frac{\rho_{m}}{\sigma_{0}}} 2 L$,  which can be interpreted as the time it takes an out-of-plane mechanical perturbation to make a roundtrip across the length $L$ of the resonator, with speed of sound $\sqrt{\frac{\sigma_{0}}{\rho_{m}}}$. As the density of the loaded region increases to the high mass load limit $\left(\rho_{m}\rightarrow \rho_{m}+\rho_{M}\right)$ the period of the fundamental mode increases, meaning $T_{\mathrm{loaded}}$ is now much longer than the roundtrip time a mechanical perturbation propagates in the unloaded region. As a result, the mode shape in the unloaded regime approximates a quasi-static displacement, with the mode shape limit being a static displacement at any moment. The second is spatial: the mode wavelength $T_{\mathrm{loaded}}\sqrt{\frac{\sigma_{0}}{\rho_{m}}}$ is much longer than the unloaded region length scale. As a result, the mode shape converges to a displacement function that doesn't curve in the unloaded region. For a string, the resulting mode shape limit is linear at the outer region, and since the inner region has to match with the outer regions at the boundaries, it implies that both the outer and inner regions approach a limit shape, which approximates a triangular shape. (Fig.~\ref{Theoretical_model}b). This leads to the splitting of the string equation as is described in Eq.~\ref{high_mass_string_equation} and Eq.~\ref{high_mass_string_equation_wave_number}, and as explained, its solution is independent of the load mass which leads to $Q$ saturation.

\par Although our analysis thus far has focused on the case of the mass loaded string, the key points of the explanation above apply more generally to highly tensioned mass loaded resonators of arbitrary geometry and imply $Q$ saturation for a high enough mass load.  Specifically, as an example we have also carried out an analysis for a circular membrane (supplementary material).

\par
\begin{figure}[t]
\centering
\includegraphics[width=0.480\textwidth]{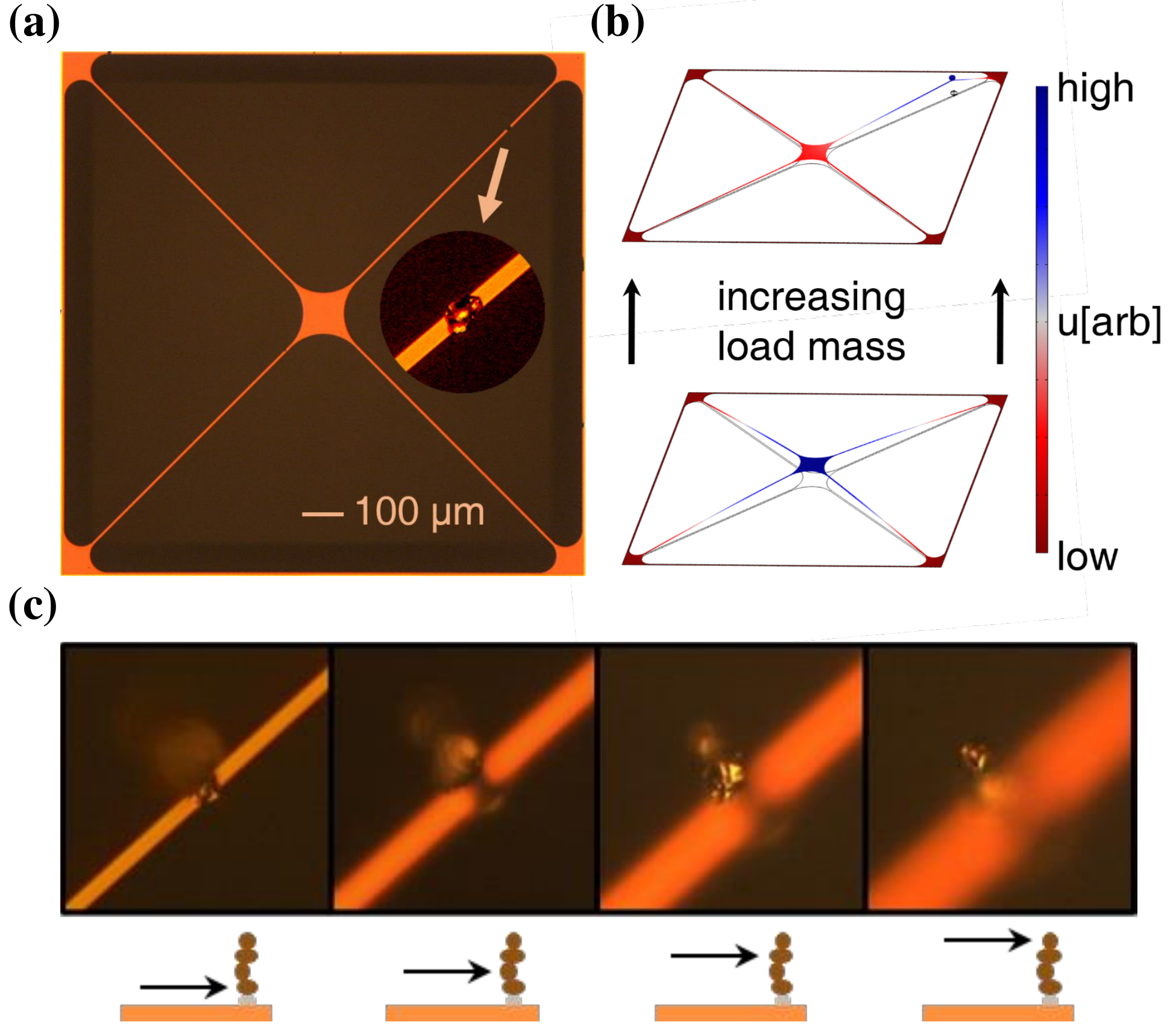}
\caption{\textbf{System for experimental validation.}\newline
\textbf{(a)} Microscope image of device: Optical microscope image shows the trampoline resonator with a magnet deposited on its tether. The inset shows a zoomed image of the magnet. \textbf{(b)} Mode shape visualization: FEA simulations show the trampoline fundamental mode shape with both a high mass load (top) and without a mass load (bottom). \textbf{(c)} Magnetic stacking: Microscope images in different planes show a stack of four magnets, going all the way from the first magnet in focus (leftmost image) to the fourth magnet in focus (rightmost image). The drawing at the bottom shows a side-view of the trampoline resonator (orange slab) with a droplet of epoxy (grey blob) and a stack of magnets (dark brown blobs). Each drawing indicates the magnet which is in focus in the corresponding microscope image (black arrow pointing at dark brown blob).}
\label{Trampoline picture with magnet}
\end{figure}

 To validate the model-predicted $Q$ saturation effect experimentally, we measured the fundamental mode $Q$, denoted hereafter as $Q_{\mathrm{fund}}$, of a high-stress SiN trampoline resonator for different load masses~\cite{norte2016mechanical,reinhardt2016ultralow}. A trampoline resonator features a wide pad for low-imprecision optical detection and the geometry is designed to reduce mechanical clamping losses, allowing access to high $Q$ modes ~\cite{norte2016mechanical, reinhardt2016ultralow}. The device we used was microfabricated with $110~\mathrm{nm}$ thick SiN with a side length of $1~\mathrm{mm}$, a tether width of $5~\mathrm{\mu m}$, and a pad area of approximately $86~\mathrm{\mu m^{2}}$ (Fig.~\ref{Trampoline picture with magnet}b). Stoichiometric SiN with a film pre-patterning  tensile stress of roughly $1~\mathrm{GPa}$ was used and an unloaded fundamental mode frequency of $\approx143~\mathrm{kHz}$ was measured. The resonator was suspended from a $385~\mathrm{\mu m}$ thick square silicon chip with a side length of $4~\mathrm{mm}$ (supplementary material).

\par In order to vary the load mass while keeping the other variables constant, we employed a ``magnetic stacking" technique. We used magnetic particles (Neo Magnequech MQFP-B+ (D50=25m)) as the load mass. The first particle was affixed to our device using ultra-violet (UV) cured epoxy (NEA 123SHGA) and is shown in Fig.~\ref{Trampoline picture with magnet}a. We then magnetized the first particle by placing the device inside a strong magnetic field. Next, a second particle was brought near the first, causing the first particle's magnetic field to induce a magnetic moment in the second particle, resulting in magnetic attraction between the particles. The device was then put under a strong magnetic field once again, this time to magnetize the second particle, and effectively make both particles a single, inseparable mass load without using additional epoxy. This procedure was repeated for multiple magnetic particles (Fig.~\ref{Trampoline picture with magnet}c) and $Q_{\mathrm{fund}}$ was measured between each addition of mass. Although we expect the epoxy to have high mechanical loss compared to SiN, by keeping its volume and three dimensional orientation constant through magnetic stacking, we were able to study how $Q_{\mathrm{fund}}$ is affected by mass loading in a systematic fashion, avoiding errors arising from multiple epoxy applications on separate devices.

\par We note that the $Q_{\mathrm{fund}}$ of trampoline resonators is sensitive to mounting due to radiation loss~\cite{norte2016mechanical, reinhardt2016ultralow}. To minimize changes in $Q_{\mathrm{fund}}$ due to varying radiation loss, we kept the mounting the same between $Q_{\mathrm{fund}}$ measurements. This was achieved by affixing our devices at one corner onto a custom-made silicon base by using silver epoxy (EPO-TEK H20E). We found that this mounting scheme consistently allowed for $Q_{\mathrm{fund}}\approx33\times10^{6}$ without a mass load. This matches the bending loss limited $Q_{\mathrm{fund}}$ obtained from FEA simulations (supplementary material), thus implying $Q_{\mathrm{rad}} \gg Q_{\mathrm{bend}}$, and therefore, $Q_{\mathrm{fund}} \approx Q_{\mathrm{bend}}$ for trampoline resonators without mass loading when mounted in this way.

\begin{figure}[t]
\centering
\includegraphics[width=0.485\textwidth]{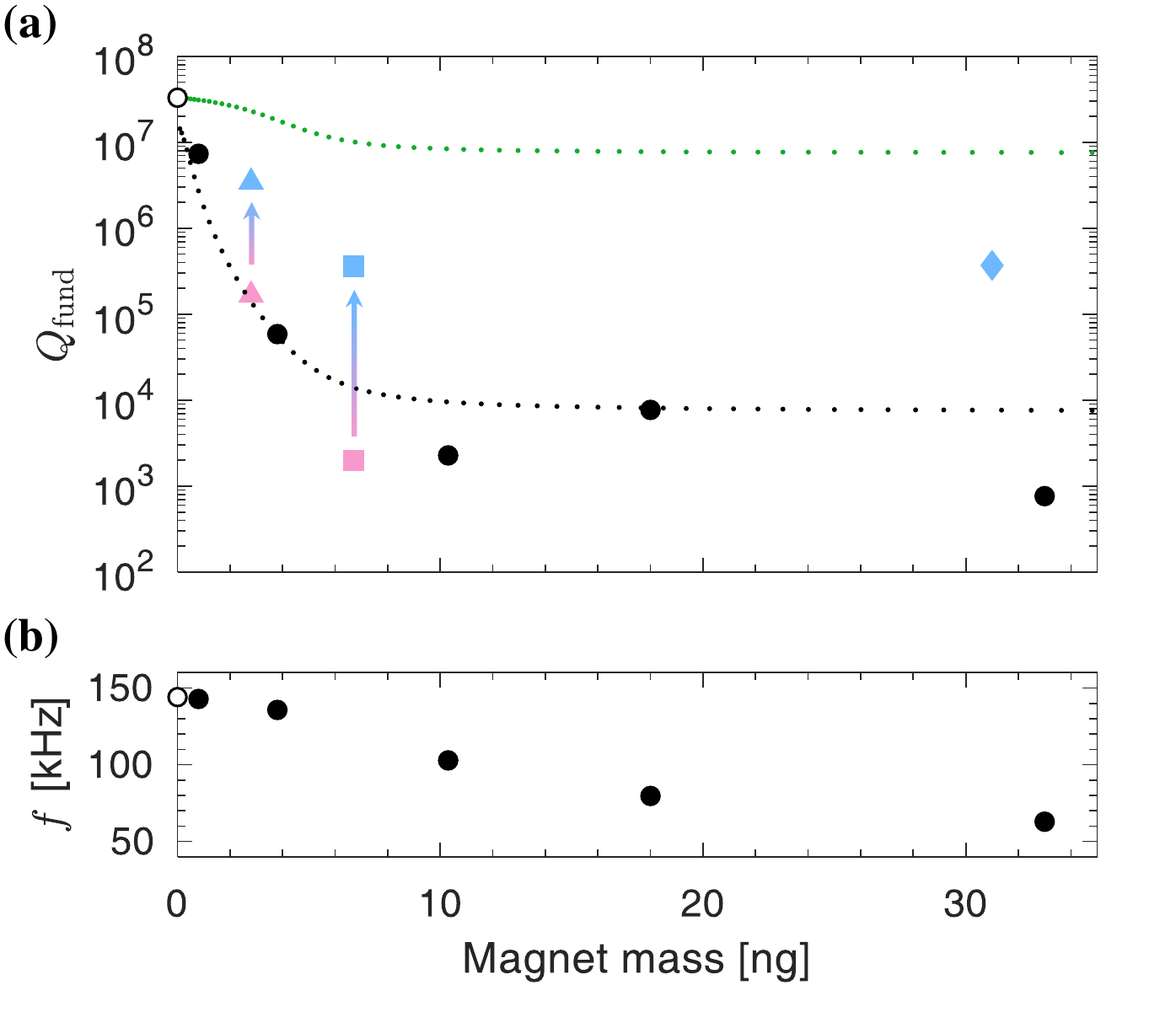}
\caption{\textbf{Experimental results and FEA simulations.}\newline
\textbf{(a)} Trampoline $Q_{\mathrm{fund}}$ as a function of magnet mass: Experimentally measured $Q_{\mathrm{fund}}$ values at 300~$\mathrm{K}$ (circles) are shown without a load mass (open black circle), and with a varying load mass (solid black circles). Corresponding FEA simulated $Q_{\mathrm{fund}}$ results at 300~$\mathrm{K}$ are shown when bending loss in the epoxy is disregarded (green dotted line), and when bending loss in the epoxy is accounted for (black dotted line). Experimentally measured $Q_{\mathrm{fund}}$ values for two other devices (triangle and square) are shown at both 300~$\mathrm{K}$ (light pink) and 8~$\mathrm{K}$ (sky blue). The first device (triangle) has mass load of 2.8~$\mathrm{ng}$ and the second device (square) has mass load of 6.7~$\mathrm{ng}$. Experimentally measured $Q_{\mathrm{fund}}$ for a third device (diamond) at 8~$\mathrm{K}$ (sky blue) with a large load mass of 31.0~$\mathrm{ng}$ is also shown. \textbf{(b)} Trampoline fundamental mode frequency as a function of magnet mass: Experimental frequency measurements at 300~$\mathrm{K}$ (circles) are shown without a load mass (open black circle) and with a varying load mass (solid black circles). Subfigures \textbf{(a)} and \textbf{(b)} share the same horizontal coordinate.}
\label{Experimental results}
\end{figure}

The key results of our experiment are shown in Fig.~\ref{Experimental results}a along with relevant FEA simulations. The open black circle and solid black circles correspond to the experimentally measured $Q_{\mathrm{fund}}$ values for a trampoline resonator at $300~\mathrm{K}$, without a load mass and with a variable load mass, respectively. They depict how $Q_{\mathrm{fund}}$ decreases as the load mass increases, but becomes distinctively insensitive at higher load mass values. This is manifested as a plateau for masses larger than $\sim10\: \mathrm{ng}$ and is indicative of $Q$ saturation. The plateau value amounts to approximately four orders of magnitude reduction in $Q_{\mathrm{fund}}$ compared to the unloaded case. To model these results, we used FEA simulations at $300~\mathrm{K}$ which took into account bending loss in both the SiN and the epoxy, while disregarding radiation loss (black dotted line). In the simulations, we scanned the epoxy length scale and loss tangent, and we chose realistic model parameters to align with our data while also setting an upper bound on $Q_{\mathrm{fund}}$ (supplementary material). We see reasonable agreement between our measured data and simulations. However, the simulated model does not fit all the data points to within the measurement noise, which we believe is because, irrespective of device mounting, at higher mass, and therefore lower frequency, the resonator is more susceptible to residual radiation loss. 

\par In order to further validate that the large reduction in $Q_{\mathrm{fund}}$ originates mainly from loss in the epoxy, rather than residual radiation loss which manifests when the load mass is large, we performed additional measurements at temperatures of $300~\mathrm{K}$ and also $8~\mathrm{K}$ (light pink and sky blue markers respectively), on two different devices (triangle and square). Each device was measured at both temperatures. The results show that the $Q_{\mathrm{fund}}$ rises by about two orders of magnitude as the resonator is cooled to cryogenic temperatures. Although SiN resonators are expected to have reduced bending loss at $8~\mathrm{K}$ compared to $300~\mathrm{K}$, empirically it is by a small factor of $\sim 3$, which cannot explain the measured difference in $Q_{\mathrm{fund}}$~\cite{reetz2019analysis}. We therefore conclude that when cold, the epoxy’s bending loss reduces significantly, which affirms that the reduction in $Q_{\mathrm{fund}}$ seen is primarily due to the lossy epoxy. 
\par A fourth device (sky blue diamond) measured at $8~\mathrm{K}$ provides evidence that a fairly high quality factor $Q_{\mathrm{fund}}\approx 3.7\times 10^{5}$ can be achieved far in the saturated regime. Previous studies have shown that using better epoxy can lead to higher quality factors~\cite{schediwy2005high}.

\par In order to set a theoretical limit on the saturated $Q_{\mathrm{fund}}$ at $300~\mathrm{K}$, we performed FEA simulations which only consider bending loss in the SiN and disregard all other forms of loss (green dotted line). The results indicate it is possible to obtain a $Q_{\mathrm{fund}}$ decrease of only a factor of $\sim 1.5$ ($Q$ as high as $10^7$) in the limit of a large load mass. The saturated $Q_{\mathrm{fund}}$ for trampolines can be further optimized by changing the location of local mass loading (supplementary material), and possibly by using trampoline resonators with carefully engineered geometries~\cite{norte2016mechanical,reinhardt2016ultralow,hoj2021ultra}.

\par In conclusion, we have analyzed the problem of tensioned mechanical resonators with a local mass load. We have shown that the added mass changes the shape of the resonator fundamental mode, and therefore, the $Q_{\mathrm{fund}}$ of the resonator. Further, we have demonstrated that for a large load mass, the mode shape becomes independent of the load mass and converges to a limit shape. This implies that $Q_{\mathrm{fund}}$ saturates in the limit of a large mass load. We have validated our model by measuring the $Q_{\mathrm{fund}}$ of a SiN trampoline resonator for different load masses. We believe that this work provides important guiding principles for the design of sensors where mass loading is needed, such as in magnetic field sensors~\cite{rugar2004single, scozzaro2016magnetic, fischer2019spin}, accelerometers~\cite{krause2012high, zhou2021broadband}, mass balances~\cite{yang2006zeptogram, ekinci2004ultrasensitive, lavrik2003femtogram}, and gravity detectors~\cite{gonzalez1994brownian, schmole2016micromechanical, liu2021gravitational, pratt2021nanoscale}. Further, the ideas presented could be extended to other high $Q$ tensioned two dimensional devices such as phononic crystals~\cite{tsaturyan2017ultracoherent, reetz2019analysis, yu2014phononic}, hierarchical structures~\cite{fedorov2020fractal, bereyhi2022hierarchical}, and machine-learning optimized resonators \cite{hoj2021ultra, shin2022spiderweb}.

\par Acknowledgements:  We thank M. Urmey for helpful comments on the manuscript.  This work was supported by funding from NSF Grant No. PHYS 1734006, Cottrell FRED Award from the Research Corporation for Science Advancement under grant 27321, and the Baur-SPIE Endowed Professor at JILA.

\bibliographystyle{prsty_all}
\bibliography{bibliography.bib}
\end{document}


\title{Supplementary Material}
\maketitle
\date{\today}

\section{Mass loaded string example}

\subsection{Finding K and frequency through wave equation}
In the limit of $l\ll L$, the frequencies of the string are well approximated by wave equation calculation. Here, we can calculate the wavenumber of the fundamental mode by solving the equation
\begin{equation}
    \frac{d^{2}u}{dx^{2}}+K\left(x\right)^{2}u = 0,
\end{equation}
where 
\begin{equation}
    K^{2}(x) = \begin{cases}
   \frac{\rho_{m}}{\sigma_{0}}\left(2\pi f\right)^{2} & 0\le x < a ,\quad b < x \le L\\
\frac{\rho_{m}+\rho_{M}}{\sigma_{0}}\left(2\pi f\right)^{2} & a\leq x \leq b 
\end{cases}  .
\end{equation}
The boundary conditions are given by
\begin{equation*}
\begin{cases}
    u(x=0)=0\\
    u(x=L)=0\\
    \frac{d^{n}u}{dx^{n}} \text{ continuous at }x=a & n=0,1\\
    \frac{d^{n}u}{dx^{n}} \text{ continuous at }x=b & n=0,1\\
\end{cases}  .
\end{equation*}

After using the first and second boundary conditions, the solution has the form:
\begin{flalign} \label{high_mass_load_solution}
   &u\left(x\right) = \nonumber\\ 
   &\begin{cases}
    A_{1}\sin\left(\frac{\alpha}{\sqrt{1+\chi}} x\right) & 0\leq x < a\\
    A_{2}\sin\left(\alpha\left( x-\frac{a+b}{2}\right)\right) + B_{2}\cos\left(\alpha\left( x-\frac{a+b}{2}\right)\right) & a\leq x \leq b \\
    A_{3}\sin\left(\frac{\alpha}{\sqrt{1+\chi}} \left(L-x\right)\right) & 0\leq x < a
    \end{cases},
\end{flalign}

This is equivalent to the more compact form in the main text. As in the main text, we defined $\alpha=\sqrt{\frac{\rho_{m}+\rho_{M}}{\sigma_{0}}\left(2\pi f\right)^{2}}$.

In order to find $\alpha$, we can write the boundary conditions matrix equation for the coefficients $A_{1},A_{2},B_{2},A_{3}$:
\begin{equation}
\left(\begin{array}{cccc}
\sin\left(\frac{\alpha}{\sqrt{1+\chi}}a\right) & -\sin\left(\alpha\left(\frac{a-b}{2}\right)\right) & -\cos\left(\alpha\left(\frac{a-b}{2}\right)\right) & 0\\
\frac{\alpha}{\sqrt{1+\chi}}\cos\left(\frac{\alpha}{\sqrt{1+\chi}}a\right) & -\alpha\cos\left(\alpha\left(\frac{a-b}{2}\right)\right) & \alpha\sin\left(\alpha\left(\frac{a-b}{2}\right)\right) & 0\\
0 & \sin\left(\alpha\left(\frac{b-a}{2}\right)\right) & \cos\left(\alpha\left(\frac{b-a}{2}\right)\right) & -\sin\left(\frac{\alpha}{\sqrt{1+\chi}}\left(L-b\right)\right)\\
0 & \alpha\cos\left(\alpha\left(\frac{b-a}{2}\right)\right) & -\alpha\sin\left(\alpha\left(\frac{b-a}{2}\right)\right) & \frac{\alpha}{\sqrt{1+\chi}}\cos\left(\frac{\alpha}{\sqrt{1+\chi}}\left(L-b\right)\right)
\end{array}\right)\left(\begin{array}{c}
A_{1}\\
A_{2}\\
B_{2}\\
A_{3}
\end{array}\right)=\left(\begin{array}{c}
0\\
0\\
0\\
0
\end{array}\right),    
\end{equation}

and find the $\alpha$ for which the determinant nulls. This results in a (typically transcendental) equation for $\alpha$, and this equation was the one used to compute the frequencies in Fig.~1c in the main text.

We can further approximate and find the large mass limit of $\alpha$. To do so, we can take $\sqrt{1+\chi}$ to be very large, such that $\sin\left(\frac{\alpha}{\sqrt{1+\chi}}x\right)\rightarrow \frac{\alpha}{\sqrt{1+\chi}}x$, $\cos\left(\frac{\alpha}{\sqrt{1+\chi}}x\right)\rightarrow 1$ for any $0\leq x \leq L$. Note that, even though $\alpha$ depends on $\sqrt{1+\chi}$, $\frac{\alpha}{\sqrt{1+\chi}}$ converges to zero as $\sqrt{1+\chi}$ increases, because as the load mass increases, the frequency of the oscillator goes to zero. Because we are looking for a zero determinant, we can multiply rows and columns by a number and the zero condition would not change. The resulting matrix is 
\begin{equation}
\left(\begin{array}{cccc}
\alpha a & -\sin\left(\alpha\left(\frac{a-b}{2}\right)\right) & -\cos\left(\alpha\left(\frac{a-b}{2}\right)\right) & 0\\
1 & -\cos\left(\alpha\left(\frac{a-b}{2}\right)\right) & \sin\left(\alpha\left(\frac{a-b}{2}\right)\right) & 0\\
0 & \sin\left(\alpha\left(\frac{b-a}{2}\right)\right) & \cos\left(\alpha\left(\frac{b-a}{2}\right)\right) & -\alpha\left(L-b\right)\\
0 & \cos\left(\alpha\left(\frac{b-a}{2}\right)\right) & -\sin\left(\alpha\left(\frac{b-a}{2}\right)\right) & 1
\end{array}\right)    
\end{equation}

and its equation for zero determinant is:
\begin{equation}
    \cos\left(\alpha\left(a-b\right)\right)\left[-\left(L-\left(b-a\right)\right)\alpha+\left(1-a\left(L-b\right)\alpha^{2}\right)\tan\left(\left(a-b\right)\alpha\right)\right]=0.
\end{equation}
which yields the two sets of solutions:
\begin{gather}
    \cos\left(\alpha\left(a-b\right)\right)=0\\
    \tan\left(\left(a-b\right)\alpha\right)=\frac{\left(L-\left(b-a\right)\right)\alpha}{\left(1-a\left(L-b\right)\alpha^{2}\right)}.
\end{gather}

The main result from this calculation is that for a large enough load mass the equation becomes load-mass independent.
Under the assumption of small mass loaded regime, i.e. $\left(b-a\right) \alpha\ll 1$, we can simplify the equation by $\tan\left((a-b)\alpha\right)\approx(a-b)\alpha$, and the fundamental mode $\alpha$ in the large mass limit is then given by 
\begin{equation}
    \alpha_{\mathrm{lim}}=\sqrt{\frac{L}{a\left(b-a\right)\left(L-b\right)}}
\end{equation}

\subsection{Outline of the full string solution}
Starting from Equation (4) and (5) in the main text, we can use a solution of the form:

\begin{flalign} \label{high_mass_load_solution}
  &u\left(x\right) = \nonumber\\ 
  &\begin{cases}
    A_{1} x+ B_{1}+ C_{1}\sinh\left(\frac{2\pi}{l} x\right) + D_{1} \cosh\left(\frac{2\pi}{l} x\right) \; 0\leq x < a\\
    A_{2}\sin\left(\beta_{t}x\right) +B_{2}\cos\left(\beta_{t}x\right) \\
    \qquad \qquad + C_{2}\sinh\left(\beta_{h}x\right) +D_{2}\cosh\left(\beta_{h}x\right)\qquad\:\: a\leq x \leq b \\
    A_{3} \left(x-L\right)+ B_{3}+ C_{3}\sinh\left(\frac{2\pi}{l} \left(x-L\right)\right) + D_{3} \cosh\left(\frac{2\pi}{l} \left(x-L\right)\right) \; a< x \leq L\\
    \end{cases},
\end{flalign}

with 

\begin{gather*}
\beta_{t}=\sqrt{\left(\frac{2\pi^{2}}{l^{2}}\right)\left(-1+\sqrt{1+4\alpha_{\mathrm{lim}}^{2}\left(\frac{l}{2\pi}\right)^{2}}\right)},\\
\beta_{h}=\sqrt{\left(\frac{2\pi^{2}}{l^{2}}\right)\left(1+\sqrt{1+4\alpha_{\mathrm{lim}}^{2}\left(\frac{l}{2\pi}\right)^{2}}\right)}.
\end{gather*}

Imposing the edge boundary conditions:
\begin{equation*}
\begin{cases}
    u(x=0)=\frac{du}{dx}(x=0)=0\\
    u(x=L)=\frac{du}{dx}(x=L)=0\\
\end{cases},
\end{equation*}
 we can simplify the solution a bit and we get:
 
\begin{flalign} \label{high_mass_load_solution}
  &u\left(x\right) = \nonumber\\ 
  &\begin{cases}
    A_{1}\left(x-\frac{l}{2\pi}\sinh\left(\frac{2\pi}{l} x\right)\right)+ B_{1}\left(1 - \cosh\left(\frac{2\pi}{l} x\right)\right) \; 0\leq x < a\\
    A_{2}\sin\left(\beta_{t}x\right) +B_{2}\cos\left(\beta_{t}x\right) \\
    \qquad \qquad + C_{2}\sinh\left(\beta_{h}x\right) +D_{2}\cosh\left(\beta_{h}x\right)\qquad\:\: a\leq x \leq b \\
    A_{3}\left(L-x  -\frac{l}{2\pi}\sinh\left(\frac{2\pi}{l} \left(L-x\right)\right)\right)\\
    \qquad \qquad+ B_{3}\left(1 - \cosh\left(\frac{2\pi}{l} \left(L-x\right)\right)\right) \qquad \quad \;\;\; b < x \leq L \\
    \end{cases}
\end{flalign}

$A_{i},B_{i},C_{i},D_{i}, i=1,2,3$ are coefficients determined from the boundary conditions
\begin{equation*}
\begin{cases}
  \frac{d^{n}u}{dx^{n}} \text{ continuous at }x=a & n=0,1,2,3\\
    \frac{d^{n}u}{dx^{n}} \text{ continuous at }x=b & n=0,1,2,3\\
\end{cases}  
\end{equation*}
and the initial condition.

\section{Mass Loaded Circular Membrane example}
\subsection{Statement of the problem}
\label{circularmemsetup}
In this section, we show that the key ideas used in the string geometry case are general, and we apply them to another geometry example - a 2D circular membrane, with a mass load at it's center.

We begin by writing the full membrane equation in 2D (corresponding to Eq.~2 in the main text):

\begin{equation}
    \left(\frac{l}{2\pi}\right)^{2}\left(\frac{\partial^{2}}{\partial x^{2}}+\frac{\partial^{2}}{\partial y^{2}}\right)^{2}u-\left(\frac{\partial^{2}}{\partial x^{2}}+\frac{\partial^{2}}{\partial y^{2}}\right)u-\frac{\rho\left(x,y\right)\left(2\pi f\right)^{2}}{\sigma_{0}}u=0.
\end{equation}

Next, we rewrite it in polar coordinates:

\begin{equation}
    \left(\frac{l}{2\pi}\right)^{2}\left(\frac{\partial^{2}}{\partial r^{2}}+\frac{1}{r}\frac{\partial}{\partial r}+\frac{1}{r^{2}}\frac{\partial^{2}}{\partial\theta^{2}}\right)^{2}u-\left(\frac{\partial^{2}}{\partial r^{2}}+\frac{1}{r}\frac{\partial}{\partial r}+\frac{1}{r^{2}}\frac{\partial^{2}}{\partial\theta^{2}}\right)u-\frac{\rho\left(r,\theta\right)\left(2\pi f\right)^{2}}{\sigma_{0}}u=0.
\end{equation}

for $0 \leq r \leq b$, and with the boundary conditions $u\left(r=b\right)=0$. Next, for simplicity, we impose a full circular symmetry:

\begin{equation}
    \left(\frac{l}{2\pi}\right)^{2}\left(\frac{\partial^{2}}{\partial r^{2}}+\frac{1}{r}\frac{\partial}{\partial r}\right)^{2}u-\left(\frac{\partial^{2}}{\partial r^{2}}+\frac{1}{r}\frac{\partial}{\partial r}\right)u-\frac{\rho\left(r\right)\left(2\pi f\right)^{2}}{\sigma_{0}}u=0,
\end{equation}

and we write the mass loading as

\begin{equation}
 \rho\left(r\right) = \begin{cases}
   \rho_{m} & 0\le r < a \\
\rho_{m}+\rho_{M} & a\leq r \leq b 
\end{cases}.
\end{equation}

\subsection{Showing high mass limit independence of loaded region wavenumber}

Like the string case, we solve for the mode's frequency by neglecting the fourth order term in the equation. this results in a Bessel equation, which yields the solution:

\begin{flalign} \label{high_mass_load_solution}
   &u\left(x\right) = \\ 
   &\begin{cases}
    A_{1} J_{0}\left(\alpha r\right) & 0\leq r < a\\
    A_{2} \left(J_{0}\left(\frac{\alpha}{\sqrt{1+\chi}} r\right) - \left(\frac{J_{0}\left(\frac{\alpha}{\sqrt{1+\chi}} b\right)}{Y_{0}\left(\frac{\alpha}{\sqrt{1+\chi}} b\right)}\right)Y_{0}\left(\frac{\alpha}{\sqrt{1+\chi}} r\right)\right) & a\leq r \leq b \\
    \end{cases},
\end{flalign}

where $J_{0}$ and $Y_{0}$ are zero order  Bessel functions of the first and second kind respectively, and like in the main text, we defined the wavenumber in the loaded regime $\alpha=2\pi f\sqrt{\frac{\rho_{m}+\rho_{M}}{\sigma_{0}}}$, using $\chi=\frac{\rho_{M}}{\rho_{m}}$, and where we already used the solution continuity at $r = 0$ and the boundary condition at $r = b$.

The remaining boundary conditions at $r = a$ can be represented as a matrix equation:

\begin{equation}
\left(\begin{array}{cc}
J_{0}\left(\alpha a\right) & -J_{0}\left(\frac{\alpha}{\sqrt{1+\chi}}a\right)+\left(\frac{J_{0}\left(\frac{\alpha}{\sqrt{1+\chi}}b\right)}{Y_{0}\left(\frac{\alpha}{\sqrt{1+\chi}}b\right)}\right)Y_{0}\left(\frac{\alpha}{\sqrt{1+\chi}}a\right)\\
-\alpha J_{1}\left(\alpha a\right) & \frac{\alpha}{\sqrt{1+\chi}}J_{1}\left(\frac{\alpha}{\sqrt{1+\chi}}a\right)-\frac{\alpha}{\sqrt{1+\chi}}\left(\frac{J_{0}\left(\frac{\alpha}{\sqrt{1+\chi}}b\right)}{Y_{0}\left(\frac{\alpha}{\sqrt{1+\chi}}b\right)}\right)Y_{1}\left(\frac{\alpha}{\sqrt{1+\chi}}a\right)
\end{array}\right)\left(\begin{array}{c}
A_{1}\\
A_{2}
\end{array}\right)=\left(\begin{array}{c}
0\\
0
\end{array}\right),
\end{equation}

which is satisfied with a non-trivial solution when the matrix determinant nulls. The equation

\begin{equation}
    \left|\left(\begin{array}{cc}
J_{0}\left(\alpha a\right) & -J_{0}\left(\frac{\alpha}{\sqrt{1+\chi}}a\right)+\left(\frac{J_{0}\left(\frac{\alpha}{\sqrt{1+\chi}}b\right)}{Y_{0}\left(\frac{\alpha}{\sqrt{1+\chi}}b\right)}\right)Y_{0}\left(\frac{\alpha}{\sqrt{1+\chi}}a\right)\\
-\alpha J_{1}\left(\alpha a\right) & \frac{\alpha}{\sqrt{1+\chi}}J_{1}\left(\frac{\alpha}{\sqrt{1+\chi}}a\right)-\frac{\alpha}{\sqrt{1+\chi}}\left(\frac{J_{0}\left(\frac{\alpha}{\sqrt{1+\chi}}b\right)}{Y_{0}\left(\frac{\alpha}{\sqrt{1+\chi}}b\right)}\right)Y_{1}\left(\frac{\alpha}{\sqrt{1+\chi}}a\right)
\end{array}\right)\right|=0,
\end{equation}

where $|\cdot|$ denotes taking determinant, is therefore an equation for $\alpha$ that satisfies the boundary conditions. \newline 
We now show that at the limit of large density ratio $\chi$, this equation becomes independent of $\chi$.
In order to do that, we use the asymptotic behavior of some of the function in the matrix element for small arguments:

\begin{equation}
    \left\{ \begin{array}{c}
J_{0}\left(\frac{\alpha}{\sqrt{1+\chi}}a\right)\rightarrow1,\qquad J_{0}\left(\frac{\alpha}{\sqrt{1+\chi}}b\right)\rightarrow1\\
Y_{0}\left(\frac{\alpha}{\sqrt{1+\chi}}a\right)\rightarrow\frac{2}{\pi}\left[\log\left(\frac{\alpha}{2\sqrt{1+\chi}}a\right)+\gamma\right],\qquad Y_{0}\left(\frac{\alpha}{\sqrt{1+\chi}}b\right)\rightarrow\frac{2}{\pi}\left[\log\left(\frac{\alpha}{2\sqrt{1+\chi}}b\right)+\gamma\right]\\
Y_{1}\left(\frac{\alpha}{\sqrt{1+\chi}}a\right)\rightarrow-\frac{1}{\pi}\left(\frac{2}{\frac{\alpha}{\sqrt{1+\chi}}a}\right)\\
J_{1}\left(\frac{\alpha}{\sqrt{1+\chi}}a\right)\rightarrow\left(\frac{\alpha}{2\sqrt{1+\chi}}a\right)
\end{array}\right.,
\end{equation}
 where $\gamma$ is a mathematical constant related to the bessel function of the second kind. Using these asymptotic approximations, we can write our equation for large $\chi$ as:
 
 \begin{equation}
     \left|\left(\begin{array}{cc}
J_{0}\left(\alpha a\right) & -\frac{\log\left(\frac{b}{a}\right)}{\log\left(\frac{\alpha}{2\sqrt{1+\chi}}b\right)+\gamma}\\
-\alpha J_{1}\left(\alpha a\right) & \left(\left(\frac{\alpha}{\sqrt{1+\chi}}\right)^{2}\frac{a}{2}+\left(\frac{\left(\frac{1}{a}\right)}{\log\left(\frac{\alpha}{2\sqrt{1+\chi}}b\right)+\gamma}\right)\right)
\end{array}\right)\right|=0.
 \end{equation}

Now, as we did in the string case, multiplying the right column of the matrix by a number does not change the solution of the equation. Therefore, we can multiply the right column by $\log\left(\frac{\alpha}{2\sqrt{1+\chi}}b\right)+\gamma$, and we get:

\begin{equation}
\left|\left(\begin{array}{cc}
J_{0}\left(\alpha a\right) & -\log\left(\frac{b}{a}\right)\\
-\alpha J_{1}\left(\alpha a\right) & \left(\frac{\alpha}{\sqrt{1+\chi}}\right)^{2}\frac{a}{2}\left(\log\left(\frac{\alpha}{2\sqrt{1+\chi}}b\right)+\gamma\right)+\left(\frac{1}{a}\right)
\end{array}\right)\right|=0.
\end{equation}
 
\begin{figure}[t]
\centering
\includegraphics[width=0.6\textwidth]{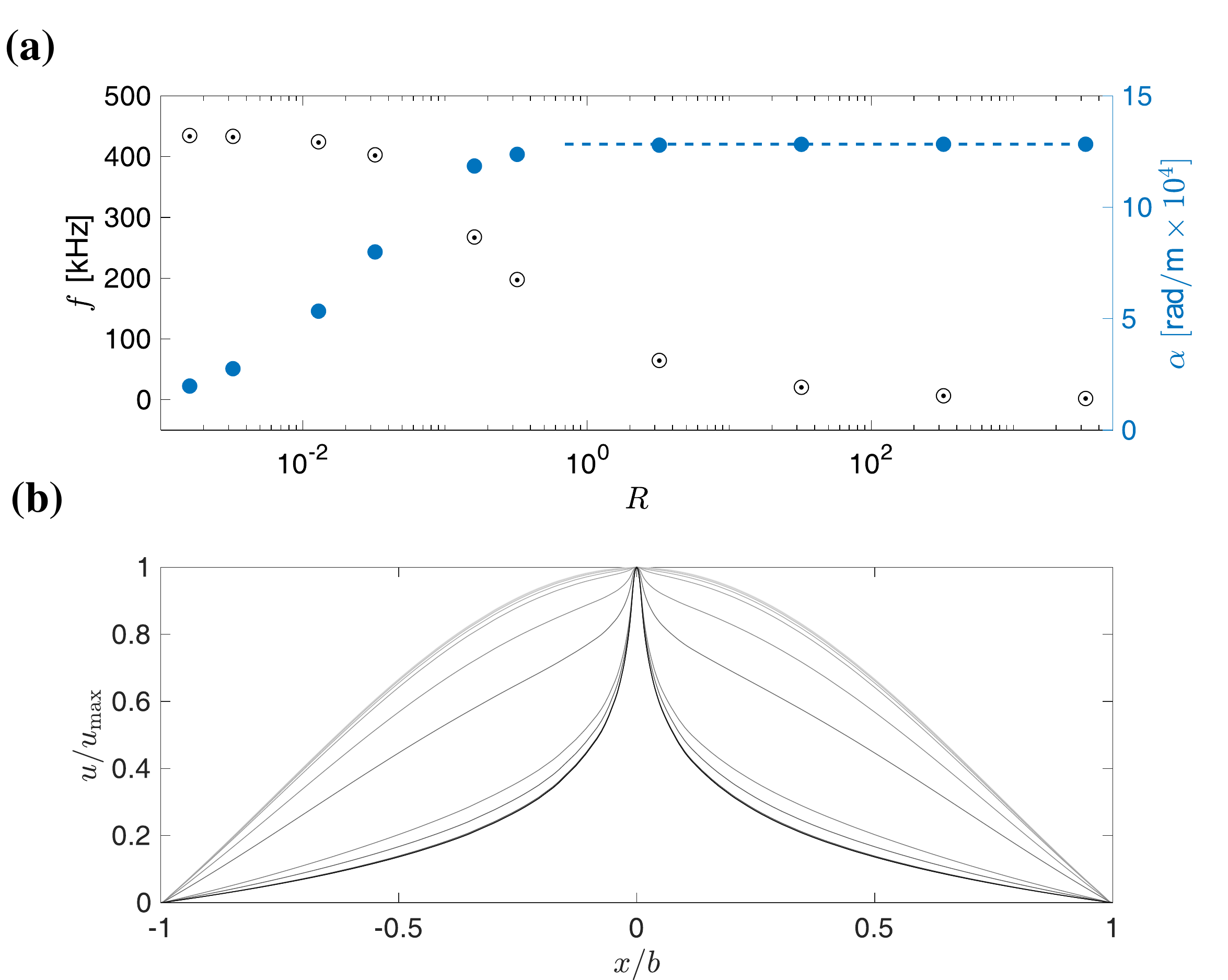}
\caption{\textbf{Mode saturation for a mass loaded circular membrane.} \textbf{(a)} Comparison between analytical model and FEA calculation for a mass-loaded circular membrane. The analytically calculated membrane frequencies (black dots) are in good agreement with the FEA simulated frequencies (empty black circles) as a function of $R$, defined as the ratio between the load mass and the total unloaded resonator mass. The calculated mass-loaded regime wavenumber $\alpha$ (blue full circles) exhibits an obvious saturation, which agrees with the limit calculated from the analytical model, $\alpha_{\mathrm{lim}}$ (dashed blue line). \textbf{(b)} Normalized mode shape visualization, for different load mass. Here, $x$ denotes the displacement from the membrane center, along a line passing through the center. Lighter shades of grey correspond to lighter load mass. The mode shapes correspond to the points in (a).} 
\label{calculation_membrane}
\end{figure}

 Now, taking the limit of large $\chi$, or $\frac{\alpha}{\sqrt{1+\chi}}\rightarrow 0$, we get the final large-mass-limit equation for $\alpha$:
 
 \begin{equation}
     \left|\left(\begin{array}{cc}
J_{0}\left(\alpha a\right) & -\log\left(\frac{b}{a}\right)\\
-\alpha J_{1}\left(\alpha a\right) & \left(\frac{1}{a}\right)
\end{array}\right)\right|=0 \Rightarrow J_{0}\left(\alpha a\right)=\alpha aJ_{1}\left(\alpha a\right)\log\left(\frac{b}{a}\right)
 \end{equation}
 
 which is independent of the mass ratio $\chi$. We can again denote the solution for the fundamental mode of this equation as $\alpha_{\mathrm{lim}}$. A comparison between the above calculation and a FEA calculation for a membrane is presented in Fig. ~\ref{calculation_membrane}. Here, we chose $b=500\: \mu\mathrm{ m}$, $a=5\: \mu\mathrm{ m}$.
 
 \subsection{Solution outline and general form}
 We can now write the large mass limit membrane equation for the fundamental mode:
 
 \begin{equation}
    \left(\frac{l}{2\pi}\right)^{2}\left(\frac{\partial^{2}}{\partial r^{2}}+\frac{1}{r}\frac{\partial}{\partial r}\right)^{2}u-\left(\frac{\partial^{2}}{\partial r^{2}}+\frac{1}{r}\frac{\partial}{\partial r}\right)u-K\left(r\right)^{2}u=0,
\end{equation}

where

\begin{equation}
 K\left(r\right)^{2} = \begin{cases}
   \alpha_{\mathrm{lim}}^{2} & 0\le r < a \\
0 & a\leq r \leq b 
\end{cases}.
\end{equation}
 
 Solving for $a\leq r \leq b$, we rewrite the equation as:
 
 \begin{equation}
     \left(\frac{\partial^{2}}{\partial r^{2}}+\frac{1}{r}\frac{\partial}{\partial r}\right)\left[\left(\frac{l}{2\pi}\right)^{2}\left(\frac{\partial^{2}}{\partial r^{2}}+\frac{1}{r}\frac{\partial}{\partial r}\right)-1\right]u=0,
 \end{equation}
 
 which can be solved by splitting it into two parts, each provides a solution with two degrees of freedom:
 
 \begin{equation}
     \left(\frac{\partial^{2}}{\partial r^{2}}+\frac{1}{r}\frac{\partial}{\partial r}\right)u=0\Rightarrow u\left(r\right)=A\ln\left(\frac{r}{B}\right),
 \end{equation}
 
 and
 
 \begin{equation}
     \left[\left(\frac{\partial^{2}}{\partial\left(\frac{2\pi}{l}r\right)^{2}}+\frac{1}{\frac{2\pi}{l}r}\frac{\partial}{\partial\left(\frac{2\pi}{l}r\right)}\right)-1\right]u=0\Rightarrow u\left(r\right)=C I_{0}\left(\frac{2\pi}{l}r\right)+D K_{0}\left(\frac{2\pi}{l}r\right),
 \end{equation}
 
 where $I_{0}\left(\cdot \right)$ and $K_{0}\left(\cdot\right)$ are the zeroth order modified Bessel functions of the first and second kind, respectively. \newline
 The solution for the $a\leq r \leq b$ regime is therefore:
 
 \begin{equation}
     u\left(r\right)=A\ln\left(\frac{r}{B}\right)+C I_{0}\left(\frac{2\pi}{l}r\right)+D K_{0}\left(\frac{2\pi}{l}r\right).
 \end{equation}
 
 Solving for the region $r<a$, we can write our equation as:
 
 \begin{equation}
     \left[\frac{1}{\eta_{h}^{2}}\left(\frac{\partial^{2}}{\partial r^{2}}+\frac{1}{r}\frac{\partial}{\partial r}\right)-1\right]\left[\frac{1}{\eta_{t}^{2}}\left(\frac{\partial^{2}}{\partial r^{2}}+\frac{1}{r}\frac{\partial}{\partial r}\right)+1\right]U=0,
 \end{equation}
 
 where
 
 \begin{gather*}
\eta_{t}=\sqrt{\left(\frac{2\pi^{2}}{l^{2}}\right)\left(-1+\sqrt{1+4\alpha_{\mathrm{lim}}^{2}\left(\frac{l}{2\pi}\right)^{2}}\right)},\\
\eta_{h}=\sqrt{\left(\frac{2\pi^{2}}{l^{2}}\right)\left(1+\sqrt{1+4\alpha_{\mathrm{lim}}^{2}\left(\frac{l}{2\pi}\right)^{2}}\right)}.
\end{gather*}

This in turn, provides a general solution in the form of:

\begin{equation}
    u\left(r\right)=A Y_{0}\left(\eta_{t}r\right)+B J_{0}\left(\eta_{t}r\right)+C I_{0}\left(\eta_{t}r\right)+D K_{0}\left(\eta_{t}r\right).
\end{equation}
 
 Taking into account continuity at the center, we conclude that the full solution has to be of the form:
 
 \begin{equation}
     u\left(r\right)=\left\{ \begin{array}{cc}
A_{1}\ln\left(\frac{r}{B_{1}}\right)+C_{1}I_{0}\left(\frac{2\pi}{l}r\right)+D_{1}K_{0}\left(\frac{2\pi}{l}r\right) & a\leq r\leq b\\
B_{2}J_{0}\left(\eta_{t}r\right)+C_{2}I_{0}\left(\eta_{t}r\right) & r<a
\end{array}\right.
 \end{equation}
 
\section{FEA Simulations}
\par We performed FEA simulations on COMSOL Multiphysics (software version 5.6) to find the expected resonator mode frequencies and $Q$ values. After setting up the computer-aided design (CAD) model of our resonator geometry along with the relevant material properties, boundary conditions, initial conditions, and mesh, we used COMSOL's Solid Mechanics Module to perform a stationary study and find the stress re-distribution. The results of this were fed into an eigenfrequency study to obtain the resonator mode shapes and frequencies. The mode shape data was then post-processed to calculate the $Q$ values. To validate the results, at the end of each simulation, we performed a mesh convergence study by systematically refining the mesh on our geometry until the resonator mode frequencies and $Q$ values changed by a negligible amount (less than $2 \%$ and $10 \%$ respectively). For all the simulations, the calculated $Q$  was based only on the bending loss inside material(s), and we ignored radiation loss. 
\par In the subsections below, we discuss the details of each individual simulation that we performed and link them to figures displayed in the main text.
\label{FEA_sims}
\subsection{Mass loaded string}
\par We simulated a mass loaded string to find the fundamental mode frequencies and $Q$ values. The results obtained from this simulation were used to create parts of Fig.~1 in the main text. The CAD model of the geometry matches what is shown in Fig.~1a of the main text. We set up the problem as a doubly-clamped string with length $L = 1~\mathrm{mm}$, width $w = 5~\mu\mathrm{m}$, thickness $h = 110~\mathrm{nm}$. For the mass loaded region we let $b-a = 10~\mu\mathrm{m}$ and placed it at the center of the string. The geometry was partitioned into the inner and outer regions (defined in the main text). The film pre-patterning stress was set to a uniform value $\sigma_p = 1~\mathrm{GPa}$ (note that this is different from the re-distributed stress $\sigma_0$). The material of the entire string was set to stoichiometric silicon nitride (SiN) with density $\rho_m = 3100~\mathrm{kg/m^3}$, Young's modulus $E = 250~\mathrm{GPa}$, and Poisson's ratio $\nu = 0.23$. However, we changed the density of the inner region to be $\rho_m + \rho_M$ where $\rho_M$ can be varied. We swept the value of $\rho_M$ over a range to change the mass load. We then computed the mode shapes and frequencies from our FEA simulation. The results were used to obtain the following quantities for different mass loads: mode shape line cuts in Fig.~1b, mode frequencies (solid black circles) in Fig.~1c, mode wavenumbers (solid blue circles) in Fig.~1c, and mode quality factors (solid black circles) in Fig.~1d. When computing the quality factors the bending loss contributions from the mode edge were ignored to focus on the main mass loading effect. The computed quality factors were normalized by the quality factor of the mode without a mass load as this gives us a dimensionless geometric parameter which does not depend on the material properties being used, and it allows us to study the effect of mass loading. As a note, when meshing the geometry, for the solutions to converge, we had to use a finer mesh around the clamped edges of the string and the mass loaded region to capture the bending accurately. Finally, we clarify that all simulations done were for the fundamental mode of the string and their purpose was to validate our analytical model.

\subsection{Mass loaded circular membrane}
We simulated a mass loaded circular membrane to find the fundamental mode shapes, wavenumbers and frequencies. The results obtained from this simulation were used to create parts of Fig.~1 in the supplementary material. The geometric setup of the problem is described above in section~\ref{circularmemsetup} of the supplementary material. We used $a = 5~\mu\mathrm{m}$ and $b = 500~\mu\mathrm{m}$ for the mass loaded membrane. The film pre-patterning stress was set to a uniform value $\sigma_p = 1~\mathrm{GPa}$ (note that this is different from the re-distributed stress $\sigma_0$). The material of the entire membrane was set to stoichiometric silicon nitride (SiN) with density $\rho_m = 3100~\mathrm{kg/m^3}$, Young's modulus $E = 250~\mathrm{GPa}$, and Poisson's ratio $\nu = 0.23$. However, we changed the density of the inner region to be $\rho_m + \rho_M$ where $\rho_M$ can be varied. We swept the value of $\rho_M$ over a range to change the mass load. We then computed the mode shapes and frequencies from our FEA simulation. The results were used to obtain the following quantities for different mass loads: mode frequencies (empty black circles) and mode inner region wavenumbers (solid blue circles) in Fig.~1a of the supplementary material, and mode shapes in Fig.~1b of the supplementary material. As a note, when meshing the geometry, for the solutions to converge, we had to use a finer mesh around the clamped edges of the membrane and the mass loaded region to capture the bending accurately. Finally, we clarify that all simulations done were for the fundamental mode of the membrane and their purpose was to validate our analytical model.

\label{mass_loaded_string}
\subsection{Trampoline resonator without mass loading}
\par We next simulated a trampoline resonator without a mass load. The results obtained from this simulation were used to create certain parts of Fig.~3 in the main text. The overall trampoline geometry used for the simulations is shown in Fig.~2a of the main text. We used experimental data for the frequency of the fundamental mode of multiple devices to set the pre-patterning stress in our COMSOL simulation. We found that $\sigma = 1.007~\mathrm{GPa}$ gave good agreement between experimental and simulated frequencies for the fundamental mode. To model the SiN we used $\rho = 3100~\mathrm{kg/m^3}$ for the density, $E = 250~\mathrm{GPa}$ for the Young's Modulus, and $\nu = 0.23$ for the Poisson's Ratio.
\par When computing the $Q$ we considered bending loss in the SiN and used Eq.~1 from the main text after some simplification. We assumed that our devices were thin-films ($h$ is small compared to the other dimensions of the resonator) which means that the bending loss is dominated by loss at the surface of the SiN, and this is indicated by $E_2$ being a function of z~\cite{tsaturyan2017ultracoherent}. Under this assumption it can be shown that Eq.~1 from the main text simplifies to the following expression~\cite{reetz2019analysis}, which can be used to compute the $Q$:
\begin{equation}
    Q_{\mathrm{bend-SiN}}(u) = \left(\frac{24(1-\nu^2)Q_{\mathrm{intrinsic-SiN}}}{h^3E}\right)\frac{W}{\int_{S}\left[\frac{\partial^{2} u}{\partial x^{2}}+\frac{\partial^{2} u}{\partial y^{2}}\right]^{2} dA}
\label{Q_bend_SiN}
\end{equation}
where $W \approx \int_{V}(\frac{\sigma_0}{2})\left[ \left(\frac{\partial u}{\partial x}\right)^{2}+\left(\frac{\partial u}{\partial y}\right)^{2}\right] dV$ is the energy stored in the mode, $S$ is the surface domain of the SiN resonator, and $dA$ is an infinitesimal surface area element. It is common practice to define $Q_{\mathrm{intrinsic}} = E/E_{2}$ where $Q_{\mathrm{intrinsic}}$ is the intrinsic quality factor of a material. The value of $Q_{\mathrm{intrinsic-SiN}}$ is variable, it depends on the specific batch of SiN used~\cite{villanueva2014evidence}.  In our case, the value of $Q_{\mathrm{intrinsic-SiN}}$ was estimated by fitting the highest experimentally measured $Q_{\mathrm{fund}}$ without a mass load, shown as an open black circle in Fig.~3a in the main text, to the simulation results. This corresponded to $Q_{\mathrm{fund}} = 33.0 \cdot 10^6$ and it gave us $Q_{\mathrm{intrinsic-SiN}} \approx 12959$ for a SiN thickness of $110~\mathrm{nm}$ at a temperature of $300~\mathrm{K}$. It should be noted that while meshing the geometry for this simulation, for the solutions to converge, we had to pay special attention to the mesh at the edges and filets of the trampoline device where the losses are significant.
\label{tramp_re_wo_mass_loading}
\subsection{Trampoline resonator with mass loading}
\par To study mass loading, we modified the simulations described in Subsec.~\ref{tramp_re_wo_mass_loading} by including a mass along with some epoxy on our trampoline resonator. We modeled the epoxy by using a small cuboid shaped slab with side length $l_g$ and thickness $t_g$ and we embedded a sphere of radius $r_{mag}$ in it to model the magnetic grain. It should be noted that we did not apply pre-patterning stress on the epoxy and the magnet. The epoxy used in the experiment was UV cured (Product ID: NEA 123SHGA) with density $\rho = 1290~\mathrm{kg/m^3}$, Young's modulus $E= 2.21~\mathrm{GPa}$, and Poisson's ratio $\nu = 0.29$. The magnets used in the experiment for mass loading were NeFeB grains (Product ID: Neo Magnequech MQFP-B+ (D50=25m)) and we modeled them using density $\rho = 7500~\mathrm{kg/m^3}$, Young's modulus $E= 160~\mathrm{GPa}$, and Poisson's ratio $\nu = 0.24$. Ideally, to change the load mass in the simulation, one would vary the sphere radius $r_{\mathrm{mag}}$. However, we chose to use a fixed value of $r_{mag}$ and scanned the density of the magnet so as to vary the load mass. This was done to avoid nonphysical torsional mode effects that show up due to mesh asymmetry when using very large values of $r_{\mathrm{mag}}$. In our simulation, we used a relatively small radius $r_{\mathrm{mag}} = 3~\mu\mathrm{ m}$ and scanned the density of the magnet grain domain over different values to simulate a variety of mass loads.
\par To compute the $Q$ of the mass loaded modes, we looked at two different scenarios: (a) disregarding loss inside the epoxy (b) including loss inside the epoxy. Additionally for both these cases, we ignored bending loss in the magnet since we expected it to be negligibly small.
\par For the first case, (a), we simply used Eq.~\ref{Q_bend_SiN} to compute the $Q$ for different mass loads. It should be noted that the stored tensile energy approximation for W mentioned earlier is challenging to calculate on COMSOL. So we used $W = 2\pi^2f^2\int_V \rho u^2 dV$ to find the maximum kinetic energy (this is equal to the stored tensile energy). This simulation gave rise to the theory points represented by the green dotted line on Fig.~3a.
\par In order to model loss inside the epoxy, (b), we used the most general formulation of bending loss which does not assume high-stress or dominant surface loss~\cite{ld1975course}. Further, it includes mechanical loss due to three dimensional deformation. With this, we obtained the following expression for the $Q$ of the epoxy domain as:
\begin{equation}
    Q_{\mathrm{bend-epoxy}}(u) = \left(\frac{2Q_{\mathrm{intrinsic-epoxy}}}{E}\right) \frac{W}{\int_{\mathrm{epoxy}} \left(\sum_{ij}(\epsilon_{ij})^2\right) dV}
\label{Q_bend_epoxy}
\end{equation}

where $\epsilon_{ij}$ is the \textit{ij}th matrix element of the three dimensional spatially dependent strain tensor. We note that here we assume an isotropic value for $E_{2}$ where $E_{2} = Q_{\mathrm{intrinsic}}/E$.
\par The $Q$ of the SiN was found using Eq.~\ref{Q_bend_SiN} as before. The $Q$ of the epoxy domain was found using Eq.~\ref{Q_bend_epoxy}. At the end, we found the overall mode $Q$ by adding the $Q$ values of the epoxy and SiN domains reciprocally. Based on surface tension arguments and additional simulations, we were able to conclude that the glue thickness $l_{t} \approx 0.8~\mu\mathr{m}$. So we scanned the values of $Q_{\mathrm{intrinsic-epoxy}}$ and the epoxy side length $l_{g}$ to find that $Q_{\mathrm{intrinsic-epoxy}} \approx 20$ and $l_{g} \approx 4~\mu\mathrm{m}$ (which are both very realistic parameters) matched our experimental data the best while also providing an upper bound on the $Q$ values. This simulation gave rise to the theory points represented by the black dotted line on Fig.~3a in the main text. It should be noted that while meshing the geometry for this simulation, for the solutions to converge, we had to pay special attention to the mesh at the edges and filets of the trampoline device, the mesh surrounding the mass loaded region, and the mesh inside the epoxy.
\label{tramp_w_mass_loading}
\section{$Q$ and Frequency Measurements}
\par To readout the mechanical motion we used an etalon interferometer with infra-red laser light of wavelength $\lambda = 1064~\mathrm{nm}$. The mode frequencies were found by identifying peaks on the measured power spectrum. To find the mode $Q$ values, we performed ringdown measurements by driving the resonator at the mode frequencies using a piezo-electric drive. The entire experiment was conducted in a high-vacuum environment ($P < 10^{-7}~\mathrm{mTorr}$) to prevent gas damping.
\section{Magnet Deposition}
\par We deposited magnets onto our trampoline resonators inside a cleanroom. This was done under a microscope using an objective lens with a magnification of 50x. Tapered glass micropipettes (Product ID: Origio MAH-SM-0) attached to a micromanipulator were used to carefully apply UV epoxy and deposit magnets (Product ID: Neo Magnequech MQFP-B+ (D50=25m)) on to our devices. The epoxy was cured using a UV flashlight. To perform magnetic stacking, we magnetized the grain(s) on our trampoline by using a strong electromagnet ($B > 1~T$). Although this field does not magnetically-saturate the grain(s), it is strong enough so as to allow us to perform magnetic stacking successfully.
\section{Microfabrication Details}
\par Devices were fabricated on a $375~\mu\mathrm{ m}$ thick, $3~\mathrm{inch}$ diameter silicon wafer with $110~\mathrm{nm}$ of stoichiometric LPCVD SiN on either side. The resonator geometry and backside windows were patterned using a direct write photolithography. Patterning of the SiN was completed via a $\mathrm{CF_4}$ reactive ion etch. The wafer was then cleaned with $\mathrm{O_2}$ plasma followed by ultrasound cleaning in an acetone bath. Additional cleaning was performed with isopropyl alcohol and water. To release the SiN structures, the window side of the wafer was etched using a 80 C KOH bath. Following wet etching, the wafer was cleaned in a Nanostrip bath, followed by a solvent clean. Devices were then left to air dry.
\section{Additional Details On Mass Loaded Trampolines}
\par In this section, we go over some additional details with regards to mass loaded trampoline resonators. To be specific, we discuss how the fundamental mode $Q$ is affected by (a) the location of the mass load on the resonator and (b) the film thickness of the resonator.
\subsection{Effect of mass loading location  on trampoline $Q$}
\par FEA simulations reveal that, for a trampoline resonator with a high mass load, the saturated $Q$ value depends on the location of mass loading. In particular, we find that the $Q$ saturates to a higher value when the mass load is on a tether rather than the trampoline pad. Further, we find the optimal location for mass loading to be the tether region near the trampoline pad. As a note, when calculating these results, we disregarded loss inside the epoxy to illustrate the best possible $Q$ values. The results are summarized in Fig.~\ref{position} of the supplementary material.
\begin{figure}[H]
\centering
    \includegraphics[width=0.6\textwidth]{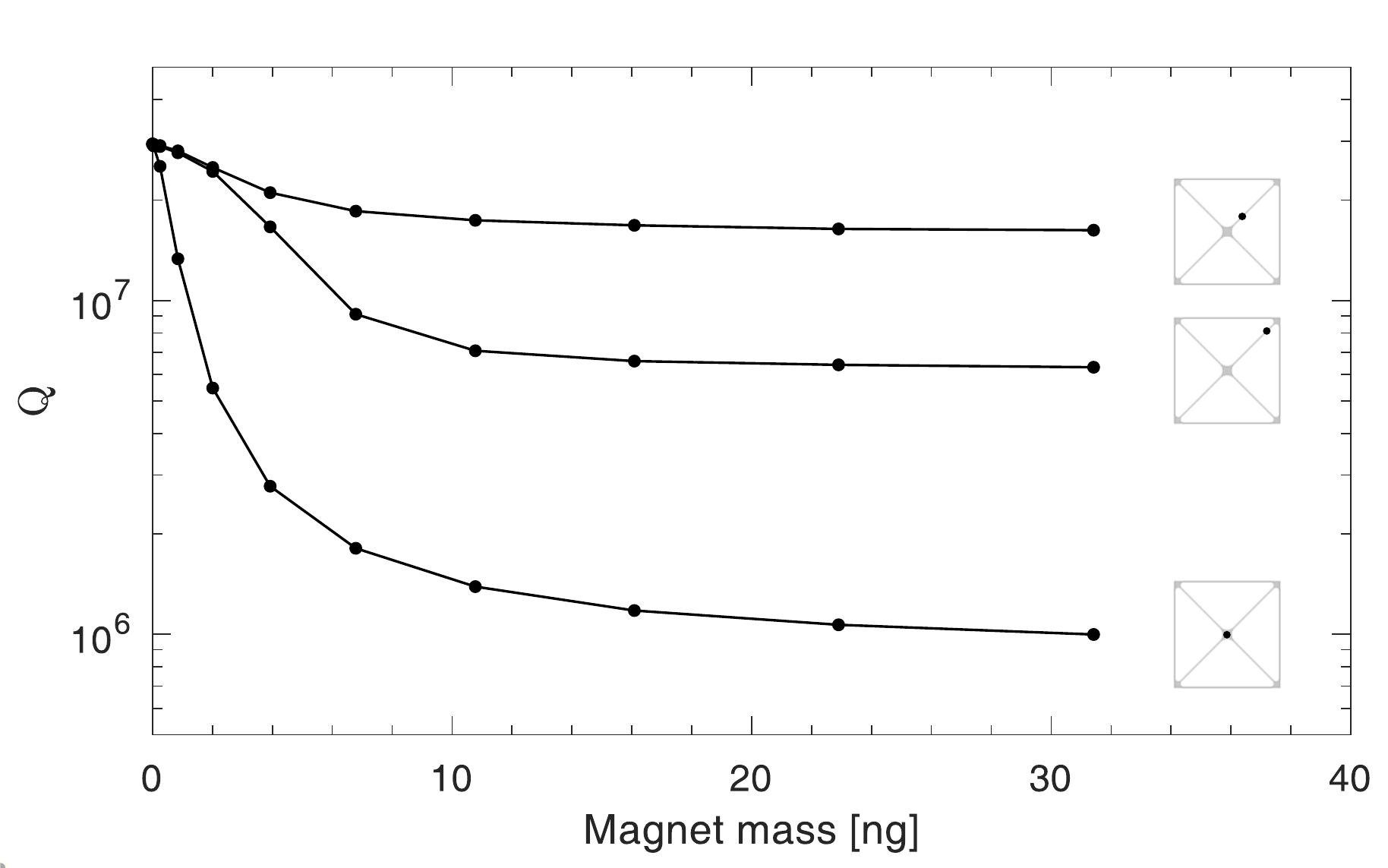}
    \caption{\textbf{Trampoline $Q$ as a function of magnet mass and location: }The plots show $Q$ for trampolines as a function of the magnet mass and location (solid black circles with black lines). The insets (small trampolines marked with a black circle) indicate the location of the mass load in each case. For each of the three curves, the solid black circles represent FEA simulation data, and we have drawn a line joining them to indicate the general trend with increasing mass load.}
    \label{position}
\end{figure}

\subsection{Effect of SiN thickness on trampoline $Q$}
It is known that for soft clamped resonators $Q \propto h^{-1}$~\cite{tsaturyan2017ultracoherent, reetz2019analysis} and because trampolines can be considered as partially soft-clamped resonators, we expect that $Q$ reduces with increasing thickness. However, when a mass load is present we find the situation is more complicated. For small masses the mode shape barely changes, and in this situation, as expected, thinner SiN offers the best $Q$ values. But as the mass increases, we observe a transition in the trend, and surprisingly, for intermediate masses, thicker SiN leads to the highest $Q$ values. However, in the limit of a large mass load, thinner SiN leads to the highest $Q$ values. But in this regime, it appears that $Q$ saturates to approximately the same value irrespective of the SiN thickness. Once again, when calculating these results, we disregarded loss inside the epoxy to illustrate the best possible $Q$ values. The results are illustrated in Fig.~\ref{thickness} of the supplementary material.

\begin{figure}[H]
\centering
    \includegraphics[width=0.6\textwidth]{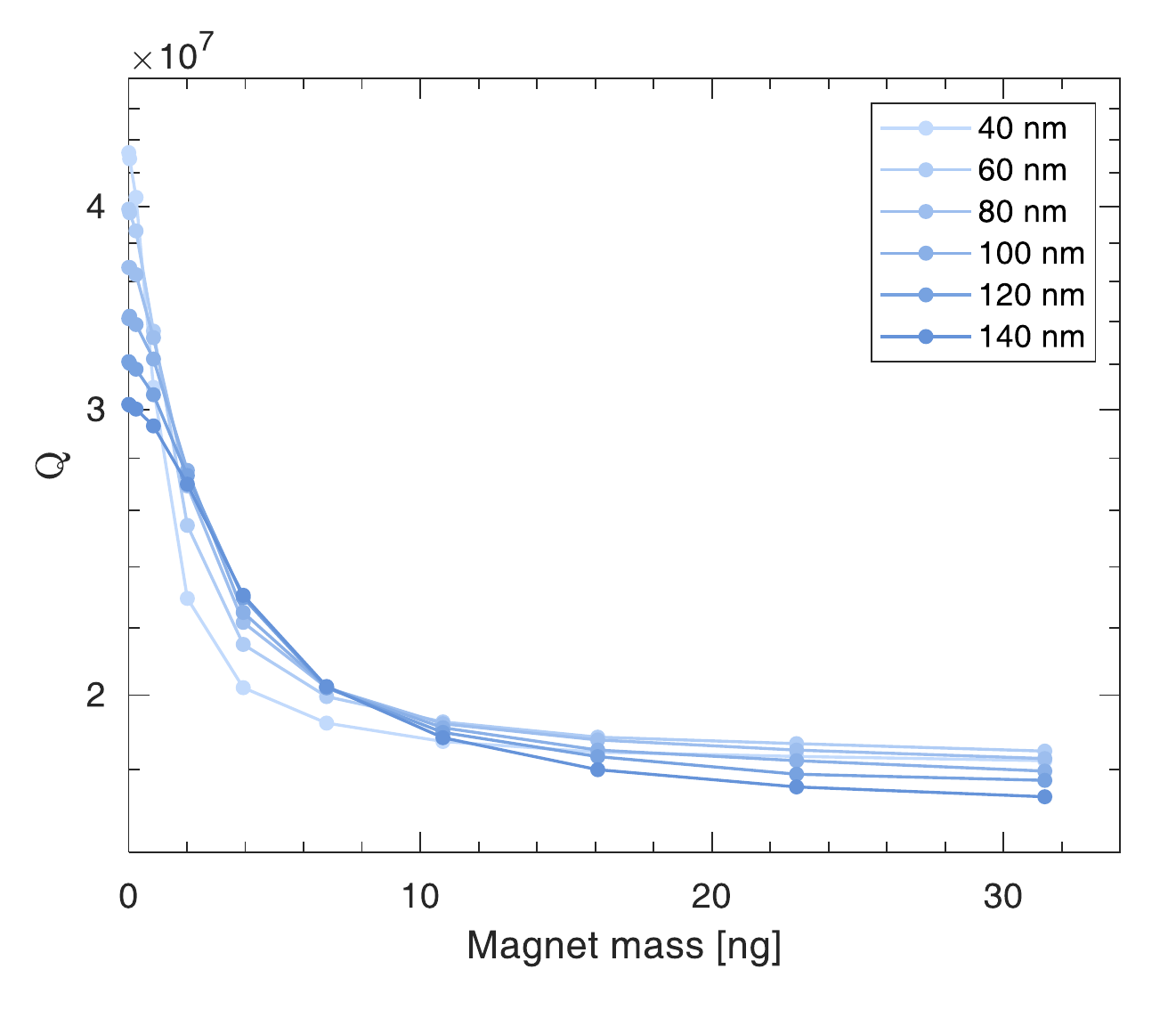}
    \caption{\textbf{$Q$ as a function of magnet mass for different SiN thicknesses: } The plots show $Q$ as a function of magnet mass for different SiN thicknesses (colored blue circles with colored blue lines). The legend (top right) indicates the nitride thickness (the lightest shade of blue corresponds to the thinnest SiN while the darkest corresponds to the thickest). For each of the curves, the solid points represent simulation data, and we have drawn a line joining them to indicate the general trend with increasing mass load.}
    \label{thickness}
\end{figure}
\bibliographystyle{prsty_all}
\bibliography{bibliography.bib}